\documentclass[prl,twocolumn,amsmath,amssymb,superscriptaddress]{revtex4-1}

\usepackage{graphicx}
\usepackage{dcolumn}
\usepackage{hhline}
\usepackage{bm,color}
\usepackage{float}
\usepackage{amsmath}
\usepackage{amsfonts}
\usepackage{amssymb}
\usepackage{epstopdf}

\begin{document}

\title{Layered palladates and their relation to nickelates and cuprates}

\author{A. S. Botana}
\affiliation{Materials Science Division, Argonne National Laboratory, Argonne, IL 60439}
\author{M. R. Norman}
\affiliation{Physical Sciences and Engineering Directorate, Argonne National Laboratory, Argonne, IL 60439}
\date{\today}

\begin{abstract}

We explore the layered palladium oxides La$_2$PdO$_4$, LaPdO$_2$ and La$_4$Pd$_3$O$_8$ via {\it ab initio} calculations.
La$_2$PdO$_4$, being low spin $d^8$, is quite different from its high spin nickel analog.
Hypothetical LaPdO$_2$, despite its $d^9$ configuration, has a paramagnetic electronic structure very different from cuprates.
On the other hand, the hypothetical trilayer compound La$_4$Pd$_3$O$_8$ ($d^{8.67}$) is
more promising in that its paramagnetic electronic structure is very similar to that of overdoped cuprates. 
But even in the $d^9$ limit (achieved by partial substitution of La with a 4+ ion), we find that an antiferromagnetic insulating state cannot be stabilized
due to the less correlated nature of Pd ions.
Therefore, this material, if it could be synthesized, would provide an ideal platform for testing the validity of magnetic theories for high temperature superconductivity.

\end{abstract}

\maketitle

\section{Introduction}

Cuprates are unique transition metal oxides in that the active transition metal $3d$ orbital ($d_{x^2-y^2}$) has a comparable energy to the $2p$ orbitals of the oxygen
ligands.  This leads to strong bonding-antibonding splitting, with a half-filled antibonding state characterizing its electronic structure \cite{pickett_cuprates}.
Chemical doping away from the stoichiometric $d^9$ configuration leads to high temperature superconductivity with a novel $d$-wave symmetry \cite{cuprates_r}.
From a materials design perspective, it is interesting to explore whether other transition metal oxides could have similar characteristics,
thus possibly leading to a new family of high temperature superconductors \cite{norman_RPP}.
In this context, layered nickelates have been an obvious target \cite{nickelate_analogs}.
The best known of the layered nickelates are the  Ruddlesden-Popper (RP) series, Ln$_{n+1}$Ni$_n$O$_{3n+1}$ (Ln being a lanthanide), composed of $n$-NiO$_2$ layers along the $c$-axis
separated by LnO spacer layers.  Their $d$ filling ranges from $d^8$ for the n=1 member to $d^7$ for the n=$\infty$ one. Single layer (n=1) Ln$_{2}$NiO$_4$ 
has been intensively studied \cite{214}. Upon Sr substitution for La, charge and spin
stripe order, a pseudogap phase, and
an insulator-to-metal transition have been reported, 
but no superconductivity has been found. 
Lower valences (higher $d$ occupations) can be achieved in RP phases by topotactic reduction. 
This was used to create infinite-layer $d^9$ LaNiO$_2$ from the cubic perovskite LaNiO$_3$ \cite{hayward}.
Although isostructural to CaCuO$_2$, its behavior is very different from that of its cuprate counterpart given the reduced $d$-$p$ hybridization 
and $d_{z^2}$ intermixing with La $d$ states \cite{lanio2_pickett}.
More promising has been the topotactic reduction of the n=3 member, Ln$_4$Ni$_3$O$_{10}$ (Ln= La, Pr).  Although La$_4$Ni$_3$O$_8$ ($d^{8.67}$)
is a charge ordered antiferromagnetic (AFM) insulator \cite{zhang},
Pr$_4$Ni$_3$O$_8$ is a metal that shares many of the characteristics of overdoped cuprates \cite{nickelates}.
However, electron doping this material has proven to be a challenge \cite{nickelate_doping}.

\begin{figure}
\includegraphics[width=0.9\columnwidth]{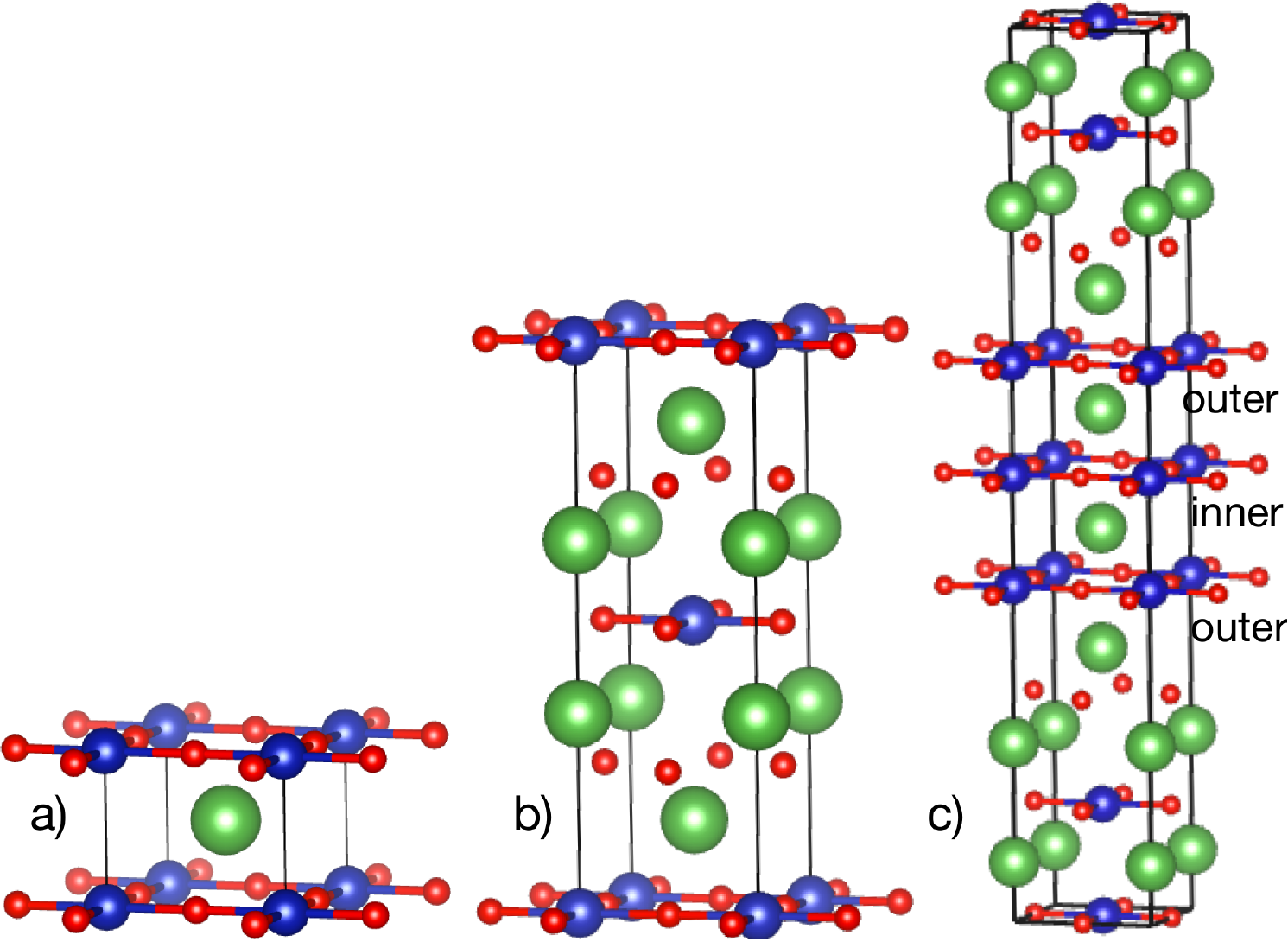}
\caption {Structure of (a) infinite layer (112), (b) single layer (214), and (c) trilayer (438) phases, with oxygen atoms in red, nickel/palladium atoms in blue, and lanthanum atoms in green. The square planar environment of the Ni/Pd ions is highlighted.}
\label{fig1}
\end{figure}

A different approach is to move down the periodic table and consider instead $4d$ analogs.  But none of the Ruddlesen-Popper phases of Ag exist.  So, we turn to Pd,
where the n=1 member (La$_2$PdO$_4$) \cite{la2pdo4} and the n=$\infty$ member (LaPdO$_3$) \cite{lapdo3} have been synthesized. Ln$_2$PdO$_4$ has been reported for a 
variety of lanthanide ions \cite{la2pdo4} and has the T$^\prime$ structure found in Nd$_2$CuO$_4$ \cite{la2pdo4_2}.
They are insulators and show diamagnetic behavior consistent with a low spin (LS) $d^8$ configuration. Electron doping by substituting La$^{3+}$ by Ce$^{4+}$ has been reported in hopes
that metallic behavior would arise \cite{suzuki}. However, the highest doping concentrations reached (20\%) are not enough to induce metallicity, 
though the resistivity is reduced.
No other Ruddlesen-Popper phases are known to exist.  This is likely because La$_4$PdO$_7$ and La$_2$Pd$_2$O$_5$ dominate the thermodynamic phase diagram \cite{jacob}.
But LaPdO$_3$ has been synthesized under pressure \cite{lapdo3}. Topotactic reduction would then give the possibility of $d^9$ Pd.  This would be significant, since 
the only formally $d^9$ Pd compounds are the delafossites such as PdCoO$_2$ \cite{mackenzie}.
But there, the Pd actually forms metallic triangular sheets and exhibits free electron-like behavior. 

Here, we use density functional theory (DFT) to study the electronic and magnetic properties of La$_2$PdO$_4$ (214 phase) as well as hypothetical LaPdO$_2$ (112 phase) and La$_4$Pd$_3$O$_8$ (438 phase),
and contrast them with their nickel analogs and with cuprates (Fig.~\ref{fig1}).  We find that the 214 and 112 phases are not promising as cuprate analogs.
However, the 438 phase has a remarkably similar paramagnetic (PM) band structure to that of overdoped cuprates.  On the other hand, we could not stabilize an antiferromagnetic insulating state in the $d^9$ limit (achieved by substituting one La$^{3+}$ per formula unit with a ${4+}$ ion). Hence, this material would be an ideal platform to test the importance of having a parent insulating phase with strong antiferromagnetic correlations for achieving high-T$_c$ superconductivity.

\section{Computational Methods}

Our electronic structure calculations were performed using the all-electron, full potential
code WIEN2k \cite{wien2k} based on the augmented plane wave
plus local orbitals (APW + lo) basis set.
Both LDA and LDA+$U$ \cite{sic} calculations were performed. For the latter, a Hund's rule $J$ is included with a typical value of 0.7 eV. The $U$ values for each calculation are specified below. For the structural optimizations, the Perdew-Burke-Ernzerhof version of the generalized
gradient approximation was used \cite{gga}.

For the calculations, we converged using R$_{mt}$K$_{max}$ = 7.0, a $k$
mesh of 14$\times$14$\times$14 for the 214 materials, 17$\times$17$\times$17 for the 438 materials, and 14$\times$14$\times$17 for the 112 ones. Muffin-tin radii of 2.38 a.u.~for La, 2.5 for Th,
1.97 a.u.~for Ni, 2.07 a.u.~for Pd, and 1.75 a.u.~for O were chosen.  Unless otherwise stated, experimental structural data were used.

\begin{figure}
\includegraphics[width=0.90\columnwidth]{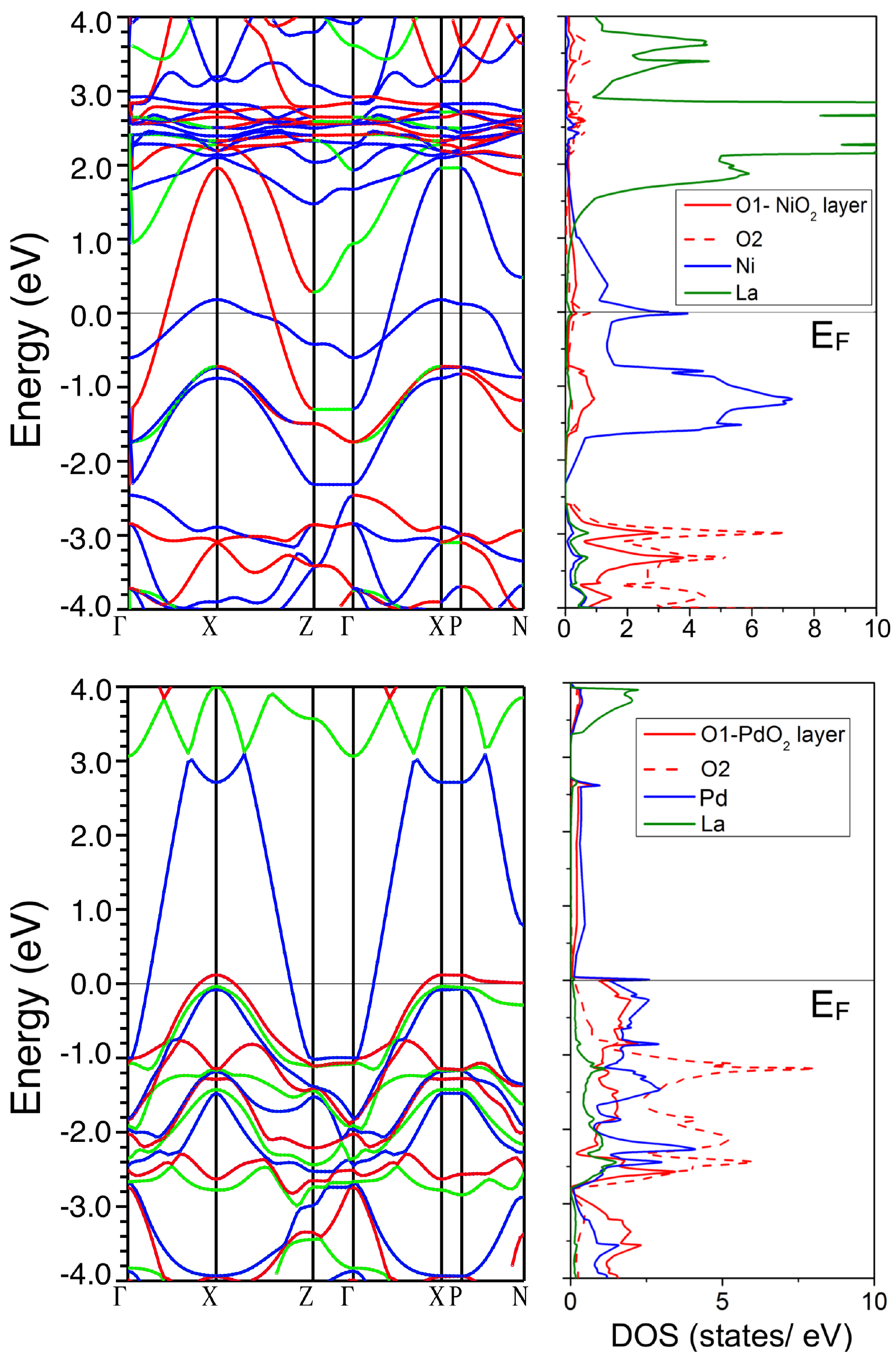}
\caption {Atom-resolved density of states (DOS) and paramagnetic GGA band structures for La$_2$NiO$_4$ (upper panels) and La$_2$PdO$_4$ (lower panels).}
\label{fig2}
\end{figure}

\section{${\rm \bf La_2PdO_4}$: Single layer}

We start with single-layer La$_2$PdO$_4$ that displays a tetragonal $I_4/mmm$ structure with
PdO$_2$ layers in which the $d^8$ ions are in a square planar environment, similar to its nickel analog
(recognizing that La$_2$NiO$_4$ has the related K$_2$NiF$_4$ structure in which the Ni ions sit in an elongated octahedral environment of oxygens \cite{la2nio4_struct}).
Let us recall some basics about the paramagnetic GGA electronic structure
of La$_2$NiO$_4$ (Fig.~\ref{fig2}).  The La-$4f$ bands do not play a role in the vicinity of the Fermi level, appearing 2.5 eV above it. The O-$2p$ bands are between -7 to -3 eV.
The Ni-$3d$ bands extend from -2.5 to 2 eV. For the Ni ions, the crystal
field splitting in a distorted octahedral environment breaks the degeneracy of the
e$_g$ states, with the d$_{x^2-y^2}$ being higher in energy with a bandwidth of $\sim$2.5 eV crossing the Fermi energy,
and a narrower $d_{z^2}$ band below it.
The on-site energy of the O-$2p$ levels is shifted down in energy by 2.5 eV with respect to the Ni-$3d$ states,
in contrast to cuprates where they are nearly degenerate.

\begin{figure}
\includegraphics[width=0.80\columnwidth]{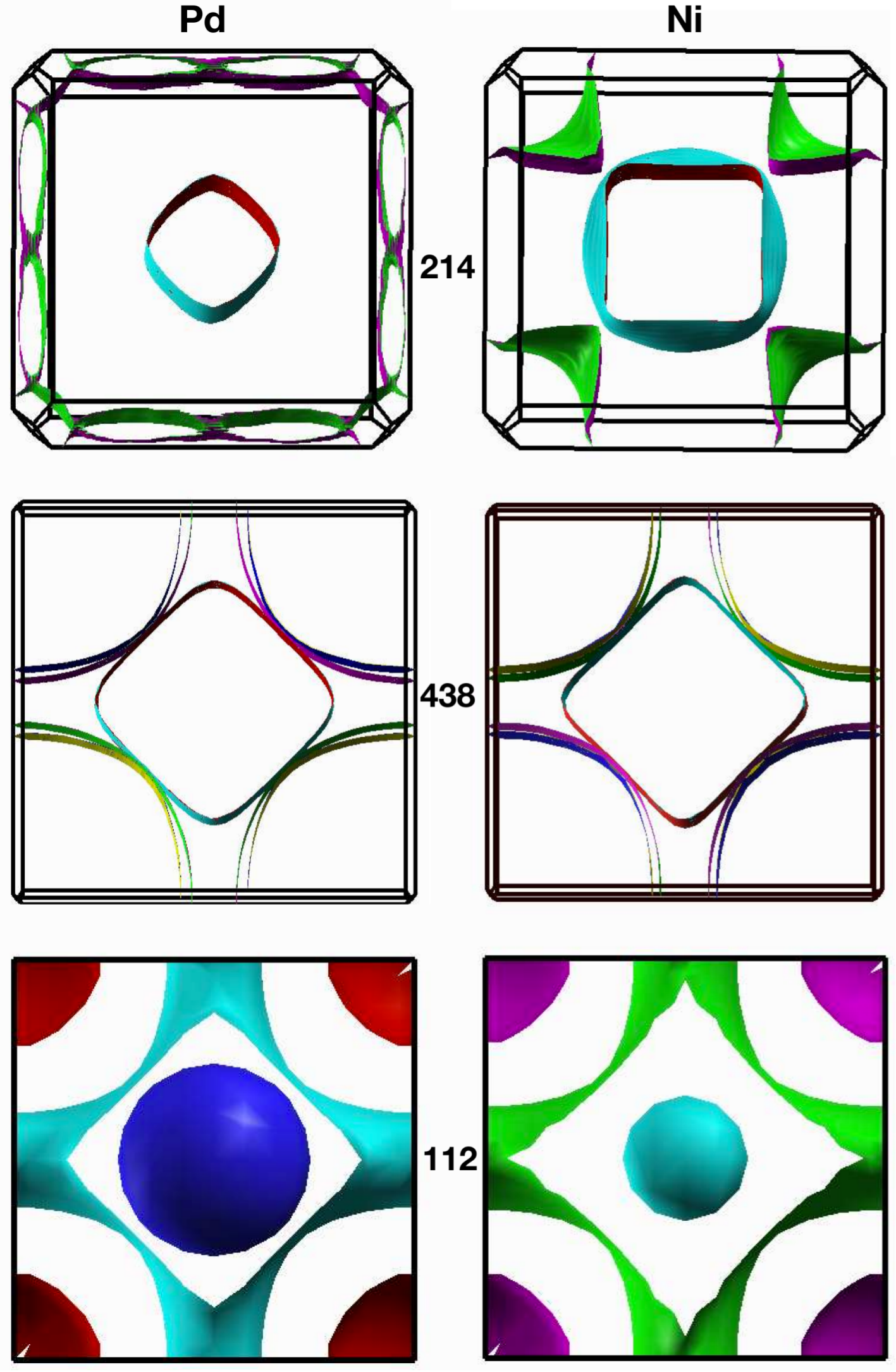}
\caption {Fermi surfaces for Pd and Ni 214, 438 and 112 phases from paramagnetic GGA calculations.}
\label{fig3}
\end{figure}

The paramagnetic GGA band structure of La$_2$PdO$_4$ is also shown in Fig.~\ref{fig2}.  The La-$4f$ bands are at higher energies. The O-$2p$ levels are shifted up in energy as well with respect to the nickel analog as can be seen in the DOS, with the hybridized O-$2p$ bands and Pd-$4d$ bands between -7 to 3 eV.
The much wider ($\sim$5 eV) $d_{x^2-y^2}$ band crossing the Fermi level reflects the stronger covalent bonding between the Pd-4$d_{x^2-y^2}$ and O-$2p$ orbitals
due to the larger radial extent of the $4d$ orbitals. For both Pd and Ni 214 phases, the involvement of $d_{z^2}$ bands gives rise to a Fermi surface different from that observed in cuprates (Fig.~\ref{fig3}).

\begin{figure}
\begin{center}
\includegraphics[width=\columnwidth]{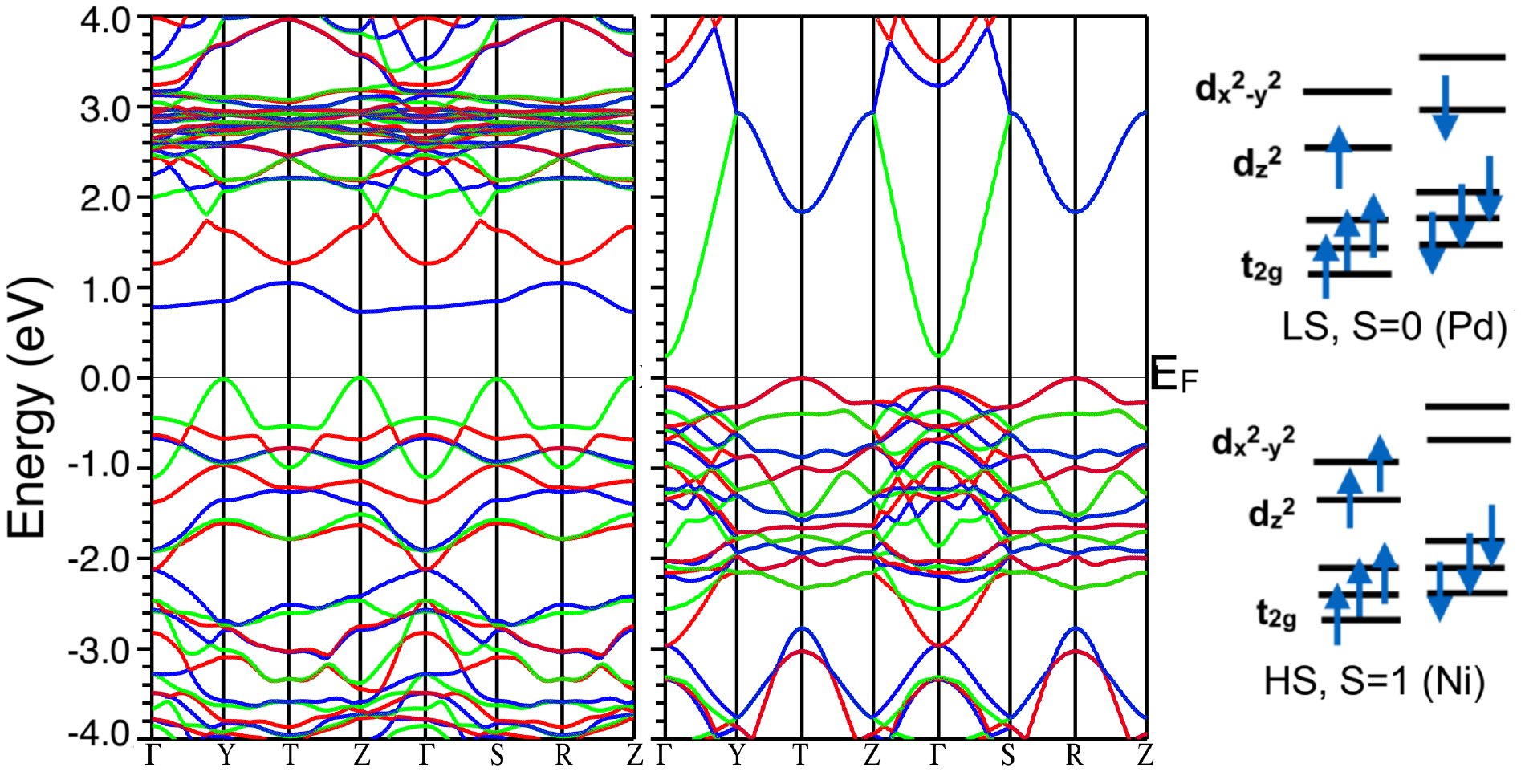}
\caption {Band structure of La$_2$NiO$_4$ (left) in an AFM state within GGA and La$_2$PdO$_4$ (middle) within LDA+$U$. Here, a $U$ of 8 eV is used to obtain an insulating state for La$_2$PdO$_4$ (though it is not magnetic).
Energy level schematic for $d^8$ high spin (HS) Ni (lower right) and low spin (LS) Pd (upper right).}
\label{fig4}
\end{center}
\end{figure}

Checkerboard antiferromagnetic (AFM) ordering even at the GGA level is sufficient to open up a gap ($\sim$ 0.8 eV) in La$_2$NiO$_4$ (Fig.~\ref{fig4}).  
This AFM state is lower in energy by 150 meV/Ni than the paramagnetic state. The significant Hund's rule
coupling stabilizes a high spin (HS) state for Ni$^{2+}$ (S = 1), with a moment of 1.31 $\mu_B$ per Ni.
The application of a Coulomb $U$ simply increases the value of the gap.

In La$_2$PdO$_4$ the situation is different. The much larger crystal field	splitting stabilizes a low spin state for Pd$^{2+}$ (S=0). This should lead to a crystal field gap between occupied d$_{z^2}$ and unoccupied d$_{x^2-y^2}$ states. But as described above, the large bandwidth closes this gap unless a Coulomb $U$ that exceeds 6 eV is applied (Fig.~\ref{fig4}).  Even then, no magnetism is found.
This value of $U$ is larger than the 3.76 eV value determined by constrained RPA calculations for Pd metal \cite{rpa}, and also larger than the 4.5 eV value estimated
from spectroscopic data on PdO \cite{uozumi}. A similar observation about $U$ has been reported for PdO \cite{bennett}, and may be connected with issues associated
with using LDA+$U$ for 4$d$ ions (it has been claimed that hybrid functionals work better for PdO \cite{derzsi} but a very large fraction of exact exchange has to be introduced).
It should be noted that the susceptibility does not give any evidence for AFM order,
simply showing T independent diamagnetic behavior, consistent with a low spin state for Pd, with a low T upturn likely due to impurities.
Regardless, it is clear that La$_2$PdO$_4$ is very different from La$_2$NiO$_4$ due to its much larger $d_{x^2-y^2}$ bandwidth and its low spin nature.

\section {${\rm \bf LaPdO_2}$: Infinite layer}

We have also studied the infinite layer hypothetical compound LaPdO$_2$. It was modeled in analogy with LaNiO$_2$ (isostructural with CaCuO$_2$) with
an assumed space group $P4/mmm$, in which both volume and $c/a$ were optimized giving rise to lattice parameters $a$=4.15 \AA~and $c$=3.47 \AA. Here, the Ni/Pd$^{1+}$ $d^9$ ions
have a square planar coordination.  

\begin{figure}
\includegraphics[width=0.9\columnwidth]{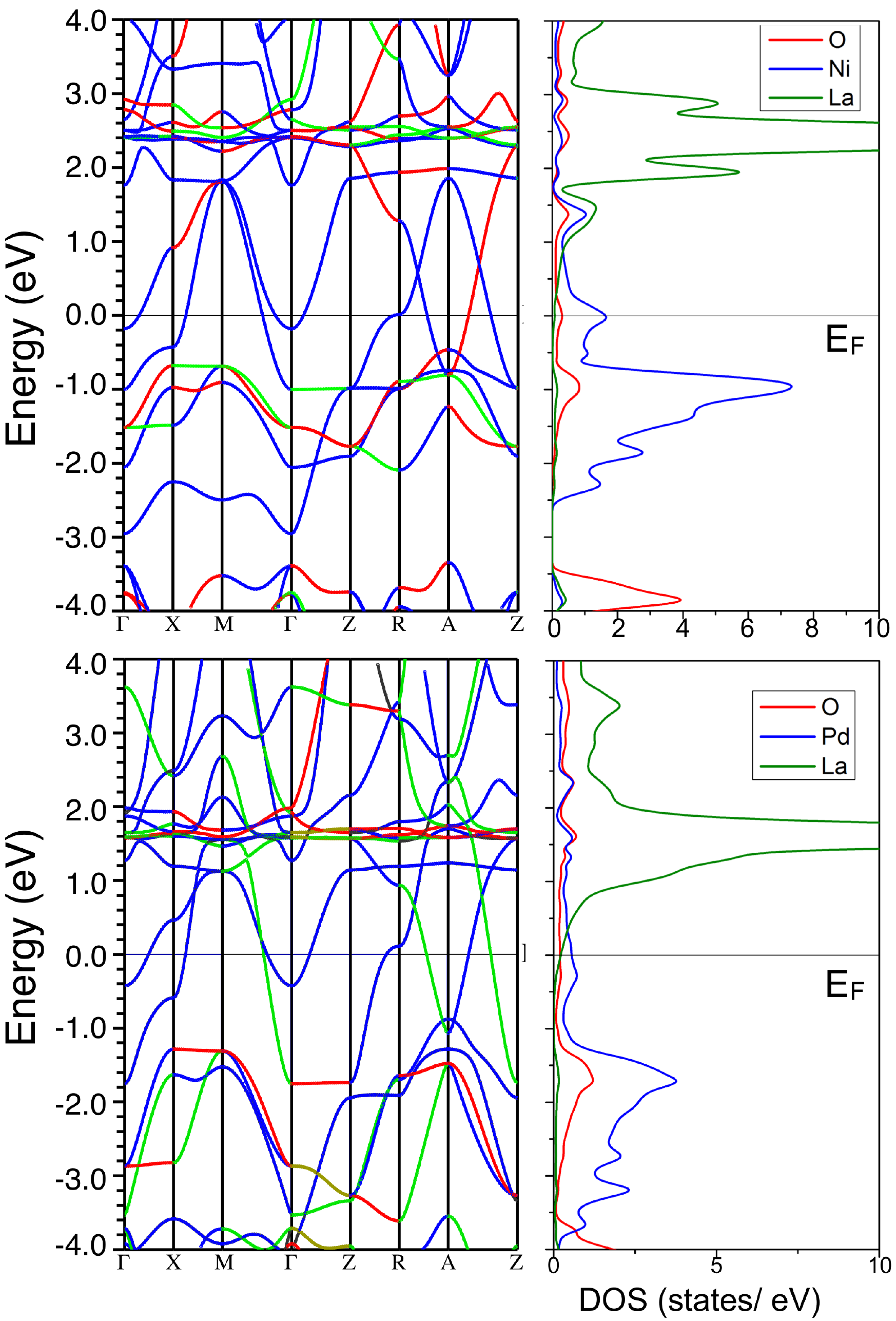}
\caption {Paramagnetic GGA band structure and atom-resolved DOS of LaPdO$_2$ (lower panels) and LaNiO$_2$ (upper panels).}
\label{fig5}
\end{figure}

We first present the GGA results (Fig.~\ref{fig5}).
For LaNiO$_2$, the La-$4f$ bands are located at 2.5 eV. The O-$2p$ bands
extend from about -8 eV to -3 eV. The Ni-$3d$ bands are distributed
from -2.5 to 2 eV. Unlike in CaCuO$_2$, there are two bands
crossing the Fermi level. One is $d_{x^2-y^2}$ in character, the other one is
a mixture of La-5$d_{z^2}$ and Ni-3$d_{z^2}$. In addition, as in La$_2$NiO$_4$, there is a smaller degree of $d-p$ hybridization as compared to cuprates.
As has been noted before, this $d^9$ metallic compound is very different from its insulating cuprate analogs \cite{lanio2_pickett}.

Is the Pd case different? In Fig.~\ref{fig5}, the band structure and DOS for LaPdO$_2$ are also shown. The La-$4f$ bands are shifted to lower energies relative to the nickelate, and hybridize with the much wider Pd-$4d_{x^2-y^2}$ band (with a bandwidth of 5 eV).  However, the electronic structure is similar in that two bands 
still cross the Fermi level: the wide Pd-4$d_{x^2-y^2}$ and the La 5$d_{z^2}$-Pd 4d$_{z^2}$ one. As in the 214 phases, the extra involvement of a $d_z^2$ band makes the Fermi surface considerably different to that of the corresponding cuprate counterpart (Fig.~\ref{fig3}).  Moreover, strong three dimensional behavior is evident
in the Fermi surface, which we find to be the case as well for the nickel analog (Fig.~\ref{fig3}), again unlike the cuprates.

A stable checkerboard AFM metallic state can be obtained for LaNiO$_2$ within GGA with a spin moment of 0.73 $\mu_B$ per Ni. This state has a lower energy by 15 meV/Ni
than that of the PM state even though there is no experimental evidence for magnetic order in this material.
In LaPdO$_2$, an AFM metallic state can also be stabilized. The derived moment on Pd$^{1+}$ within GGA is lower than in the nickelate case (0.22 $\mu_B$). We find that the PM state is more stable than the AFM one by 4 meV/Pd.

Given that substitution of Ni by Pd was not promising in terms of eliminating the $d_{z^2}$ contribution around E$_F$, we turn back to LaNiO$_2$ where one could attempt to substitute La by another 3+ ion to eliminate the 
problematic La-5d$_{z^2}$ states from the vicinity of the Fermi energy. One approach might be to substitute La by the smaller Y.
The $P4/mmm$ structure was assumed once again with the lattice constants being optimized.
The band structure of hypothetical YNiO$_2$ is quite similar to that of its La counterpart with Y-4$d_{z^2}$ states still crossing E$_F$, giving rise to even larger Fermi surface pockets than in LaNiO$_2$.  We also tried using Tl (with its filled d shell) in place of Y, but two bands still cross E$_F$.

\section{${\rm \bf La_4Pd_3O_8}$: Trilayer}

\begin{figure}
\includegraphics[width=0.92\columnwidth]{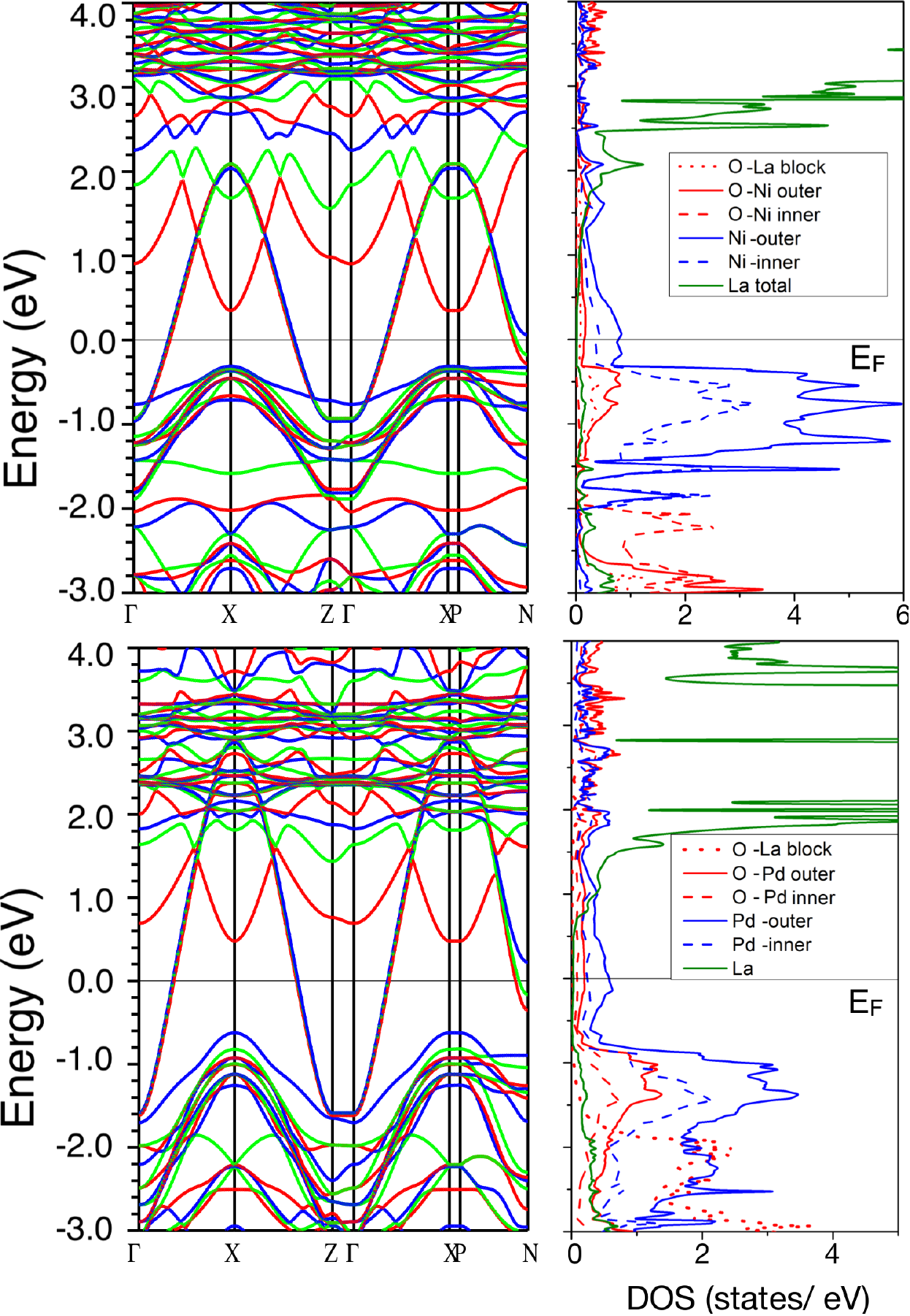}
\caption {Paramagnetic GGA band structure and atom-resolved DOS of La$_4$Pd$_3$O$_8$ (lower panels) and La$_4$Ni$_3$O$_8$ (upper panels).}
\label{fig6}
\end{figure}

After ruling out the 214 and 112 layered palladates as cuprate analogs, we turn our attention to the 438 phases.
Ln$_4$Ni$_3$O$_8$ (Ln= La, Pr) has a tetragonal $I_4/mmm$ structure in which the Ni is in a square planar environment.
Although the La438 phase is known to be a charge-ordered insulator \cite{zhang}, Pr438 is metallic \cite{nickelates}.
Therefore, we focus on the paramagnetic electronic structure of La438 (Fig.~\ref{fig6}).
Again, the La-$4f$ bands are located more than 3 eV above the Fermi level. The O-$2p$ bands
extend from about -5 eV to 2 eV. The Ni-$3d$ bands are distributed
from -2 to 2 eV.  A single band per Ni with d$_{x^2-y^2}$ character and a 3 eV bandwidth due to large $d$-$p$ hybridization crosses the Fermi level. 

Although higher-order RP phases are not known for the palladates, it is possible that they could be stabilized under high oxygen partial pressures.
If the n=3 phase could be stabilized, then the resulting 438 phase could conceivably be made by reduction.  To explore this possibility,
the Pd analog was modeled from the Ni 438 structure by performing an optimization of both volume and $c/a$ giving rise to lattice parameters $a$=4.17 \AA~and $c$=26.31 \AA.
We find that the the electronic structure of the palladate is very similar to that of the nickelate (Fig.~\ref{fig6}), though with a larger bandwidth (4.4 eV) due to
increased $d$-$p$ hybridization. We also find that the Fermi surfaces of Pd and Ni-438 phases are very similar and cuprate-like (Fig.~\ref{fig3}).
Note the presence of two large hole-like cylinders and one electron-like cylinder, indicating that the doping level of the inner plane differs from that
of the two outer planes (Fig.~\ref{fig1}), as has been inferred in trilayer (and higher-layer) cuprates by NMR \cite{nmr}.
Given the similarity of the Fermi surfaces, one might then expect that an electron-doped version of this layered palladate might be a promising candidate
for high temperature superconductivity \cite{eriksson}.
This is particularly relevant, given that substituting La by a smaller 4+ ion in the nickel analog has proven to be difficult; the larger size of Pd versus Ni may make this easier for the Pd case. 

To explore this further, we investigated whether a magnetic state could be obtained.  Previously, we had been able to stabilize magnetic solutions for
La$_4$Ni$_3$O$_8$ and Pr$_4$Ni$_3$O$_8$ if we include an on-site Coulomb $U$ \cite{nickelates,antia1,antia2}. In the 438 phases, the average Ni/Pd valence is +1.33 which leaves the e$_g$ orbitals with 2.67 electrons per Ni/Pd on average. If the Hund's rule coupling is larger than the splitting between the two $e_g$ orbitals, the transition metal ion would be in a HS state, in the opposite situation, in a LS state. 
If the latter occurs, all the $z^2$ bands (majority and minority spin) and 2/3 of a majority spin $x^2-y^2$ band will be occupied. Due to this partially filled band crossing the Fermi level, the 438 materials in a LS state are metallic, with all unoccupied states having $x^2-y^2$ parentage. The resulting magnetic order in this case is ferromagnetic (FM). This is the magnetic ground state we find in Pr$_4$Ni$_3$O$_8$ and also in hypothetical La$_4$Pd$_3$O$_8$. The electronic structure of the majority spin channel in this FM-LS state is analogous to that of the PM state shown in Fig.~\ref{fig6}. As in the 112 case, the derived magnetic moments are smaller for Pd (0.50 $\mu_B$) than for Ni (0.70 $\mu_B$) at the same $U$ value (3.5 eV).
For the palladate, given the large crystal field splitting, a HS state could not be stabilized (unlike for the nickelate).

To investigate further, we explored the effects of electron doping the 438 palladate.
For ThLa$_3$Pd$_3$O$_8$, Pd should have a $d^9$ configuration, the same as in parent cuprates. We chose Th over Ce to avoid dealing with $f$ states and because Ce has two plausible oxidation states (3+ and 4+). Here, an AFM state could be stabilized within LDA+$U$, but a gap cannot be opened regardless of the $U$ value due to the much stronger hybridization of Pd-$d$ and Th-$d$ states as compared to the nickelate \cite{antia2}. Though this metallic AFM state is more stable than a FM one by 27 meV/Pd  for $U$=3.5 eV, this energy difference is one order of magnitude smaller than in the nickelate.  The derived magnetic moment per Pd is 0.50 $\mu_B$ for this $U$ value.
Regardless, this AFM metallic solution differs from the AFM insulating solution found in the $3d$ nickelates and cuprates.

La$_4$Pd$_3$O$_8$ is then a cuprate analog in that it has a quasi-2D crystal structure, with strong $p$-$d$ hybridization, a single band of $d_{x^2-y^2}$ character per Pd crossing the Fermi level, and a similar Fermi surface topology \cite{eriksson,fs_wep}. Yet, the material is different in that a parent insulating antiferromagnetic phase is predicted to be absent. Hence, this material would be a good platform to investigate the importance of having an insulating phase with strong antiferromagnetic correlations for high-T$_c$ superconductivity, which would test whether superexchange is playing a key role as advocated by a number of authors \cite{pwa,lee,scalapino}.

\section{summary}

To summarize, the substitution of Ni with Pd in 214-single and 112-infinite layer compounds (aside from their formal structural similarities) induces drastic changes in the
electronic structure and magnetic properties, mainly due to the much larger crystal field splitting and bandwidth of the Pd compounds.
Therefore, we argue that the route to search for new high temperature superconductors by doping these layered palladates is not a promising one. However, the 
hypothetical 438 material shares many cuprate-like features in terms of its paramagnetic electronic structure and fermiology, yet it is different in that a parent insulating antiferromagnetic phase could not be stabilized.
Therefore, if this material could be synthesized, it would be an ideal platform to test the importance of antiferromagnetic correlations for high temperature superconductivity. 

\section{acknowledgments}

This work was supported by the Materials Sciences and Engineering
Division, Basic Energy Sciences, Office of Science, US DOE. 
We acknowledge the computing resources provided on Blues, a high-performance computing clusters operated by the Laboratory
Computing Resource Center at Argonne National Laboratory.


\begin{thebibliography}{99}

\bibitem{pickett_cuprates}
W. E. Pickett, Rev. Mod. Phys. {\bf 61}, 433 (1989).
\bibitem{cuprates_r}
B. Keimer, S. A. Kivelson, M. R. Norman, S. Uchida and J. Zaanen, Nature {\bf 518}, 179 (2015).
\bibitem{norman_RPP}
M. R. Norman, Rep. Prog. Phys. {\bf 79}, 074502 (2016).
\bibitem{nickelate_analogs}
V. I. Anisimov, D. Bukhvalov and T. M. Rice, Phys. Rev. B {\bf 59}, 7901 (1999).
\bibitem{214}
M. Uchida, K. Ishizaka, P. Hansmann, Y. Kaneko, Y. Ishida, X. Yang, R. Kumai, A. Toschi, Y. Onose, R. Arita,
K. Held, O. K. Andersen, S. Shin and Y. Tokura, Phys. Rev. Lett. {\bf 106}, 027001 (2011).
\bibitem{hayward}
M. A. Hayward, M. A. Green, M. J. Rosseinsky and J. Sloan, J. Amer. Chem. Soc. {\bf 121}, 8843 (1999).
\bibitem{lanio2_pickett}
K.-W. Lee and W. E. Pickett, Phys. Rev. B {\bf 70}, 165109 (2004).
\bibitem{zhang}
J. Zhang, Y.-S. Chen, D. Phelan, H. Zheng, M. R. Norman and J. F. Mitchell, Proc. Natl. Acad. Sci. {\bf 113}, 8945 (2016).
\bibitem{nickelates}
J. Zhang, A. S. Botana, J. W. Freeland, D. Phelan, H. Zheng, V. Pardo, M. R. Norman and J. F. Mitchell,
Nature Physics {\bf 13}, 864 (2017).
\bibitem{nickelate_doping}
Junjie Zhang and John Mitchell, private communication.
\bibitem{la2pdo4}
S. Shibasaki and I. Terasaki, J. Phys. Soc. Jpn. {\bf 75}, 024705 (2006).
\bibitem{lapdo3}
S.-J. Kim, S. Lemaux, G. Demazeau, J.-Y. Kim and J.-H. Choy, J. Amer. Chem. Soc. {\bf 123}, 10413 (2001).
\bibitem{la2pdo4_2}
J.P. Attfield and G. Ferey, J. Solid State Chem. {\bf 80}, 286 (1989).
\bibitem{suzuki}
S. Suzuki, K. Kawashima, W. Ito, S. Igarashi, M. Yoshikawa and J. Akimitsu, JPS Conf. Proc. {\bf 3}, 017028 (2014).
\bibitem{jacob}
K. T. Jacob, K. T. Lwin and Y. Waseda, Solid State Sci. {\bf 4}, 205 (2002).
\bibitem{mackenzie}
A. P. Mackenzie, Rep. Prog. Phys. {\bf 80}, 032501 (2017).
\bibitem{wien2k}
P. Blaha, K. Schwarz, G. K. H. Madsen, D. Kvasnicka and J. Luitz,
{\it WIEN2k, An Augmented Plane Wave Plus Local Orbitals
  Program for Calculating Crystal Properties}, Vienna University of Technology, Austria (2001).
\bibitem{sic}
A. Lichtenstein, V. Anisimov and J. Zaanen, Phys. Rev. B {\bf 52}, R5467 (1995).
\bibitem{gga}
J. P. Perdew, K. Burke and M. Ernzerhof, Phys. Rev. Lett. {\bf 77}, 3865 (1996).
\bibitem{la2nio4_struct}
 S. J. Skinner, Solid State Sci. {\bf 5}, 419 (2003).
\bibitem{rpa}
E. Sasioglu, C. Friedrich and S. Blugel, Phys. Rev. B {\bf 83}, 121101 (2011).
\bibitem{uozumi}
T. Uozumi, T. Okane, K. Yoshii, T. A. Sasaki and A. Kotani, J. Phys. Soc. Jpn. {\bf 69}, 1226 (2000).
\bibitem{bennett}
J. W. Bennett, I. Grinberg, P. K. Davies and A. M. Rappe, Phys. Rev. B {\bf 82}, 184106 (2010).
\bibitem{derzsi}
M. Derzsi, P. Piekarz and W. Grochala, Phys. Rev. Lett. {\bf 113}, 025505 (2014).
\bibitem{nmr}
H. Mukuda, S. Shimizu, A. Iyo and Y. Kitaoka, J. Phys. Soc. Jpn. {\bf 81}, 011008 (2012).
\bibitem{eriksson}
M. Klintenberg and O. Eriksson, Comp. Mater. Sci. {\bf 67}, 282 (2013).
\bibitem{antia1}
A. S. Botana, V. Pardo, W. E. Pickett and M. R. Norman, Phys. Rev. B {\bf 94}, 081105(R) (2016).
\bibitem{antia2}
A. S. Botana, V. Pardo and M. R. Norman, Phys. Rev. Materials {\bf 1}, 021801 (2017).
\bibitem{fs_wep}
W. E. Pickett, H. Krakauer, R. E. Cohen, and D. J. Singh, Science {\bf 255}, 46 (1992).
\bibitem{pwa}
P. W. Anderson, Science {\bf 235}, 1196 (1987).
\bibitem{lee}
P. A. Lee, N. Nagaosa and X.-G. Wen, Rev. Mod. Phys. {\bf 78}, 17 (2006).
\bibitem{scalapino}
D. J. Scalapino, Rev. Mod. Phys. {\bf 84}, 1383 (2012).

\end{thebibliography}
\end{document}